\documentclass{webofc}
\usepackage[varg]{txfonts}   

\usepackage{xcolor}

\usepackage{hyperref}
\hypersetup{
    colorlinks=false,
    linkcolor=black,
    filecolor=magenta,      
    urlcolor=blue,
}

\begin{document}

\title{Measurement of muon flux variations with scintillators in Mexico City and their correlation to solar activity}

\author{
\firstname{Diego} \lastname{Martínez Montiel}\inst{1}\fnsep\thanks{\email{diego_montiel@ciencias.unam.mx}} 
\and \firstname{Leonid} \lastname{Serkin}\inst{2}\fnsep\thanks{\email{leonid.serkin@correo.nucleares.unam.mx}}
\and \firstname{Guy} \lastname{Paic}\inst{2}
\and \firstname{Miguel Enrique} \lastname{Patiño Salazar}\inst{2}
\and \firstname{Jaime Octavio} \lastname{Guerra Pulido}\inst{2}
\and \firstname{Carlos Rafael} \lastname{Vázquez Villamar}\inst{2}
}

\institute
{Facultad de Ciencias, Universidad Nacional Autónoma de México, Circuito Exterior s/n, Ciudad de México 04510, México. 
\and
Instituto de Ciencias Nucleares, Universidad Nacional Autónoma de México, Apartado Postal 70-543, Ciudad de México 04510, México.
}

\abstract{%
In this work we report the variations of the atmospheric muons flux in Mexico City. The measurements were performed from September to December 2023 at at the Institute of Nuclear Sciences of UNAM, located at coordinates 19.32$^{\circ}$N 99.18$^{\circ}$W, with an altitude of 2274 meters above see level and a cutoff rigidity of 8.24 GV. The experimental setup consists of a pair of scintillators, with dimensions of 50cm x 1m. To compensate for fluctuations in air pressure, we calculate the barometric coefficient using data, which is determined to be -0.21\%/mb. The observed muon flux corresponds to $98.3 \pm 2.6$ m$^{-2}$ s$^{-1}$ sr$^{-1}$, and is consistent with other measurements at the same location. We have identified three variations in the pressure-corrected muon flux that exhibit a pattern similar to those observed by a neutron cosmic ray detector at UNAM. Through an extensive analysis of the publicly available geomagnetic data, we have correlated these variations with a phenomenon known as a Forbush decrease.
}
\maketitle
\section{Introduction} 
\label{intro} 
It is generally accepted that the primary flux of cosmic rays is predominantly isotropic, arriving at Earth from all directions in space with roughly equal intensity~\cite{Int1}. It comprises approximately 89\% protons, 9\% helium nuclei, and the remaining 2\% is distributed among electrons, positrons, and heavy nuclei~\cite{PGD}. When cosmic rays encounter Earth's atmosphere, they interact with the atoms that compose it, generating secondary radiation. Muons are the products of the decay of charged pions produced in such interactions, and due to the relativistic dilatation of time, constitute the majority of the cosmic radiation incident on the Earth's surface. The average energy of muons at sea level is around 4 GeV, and the muon flux at sea level is 1 muon per cm$^2$ per second. 

\medskip \noindent 
The Sun plays a significant role in modulating cosmic radiation, leading to fluctuations known as solar modulation. Periods of high solar activity, known as geomagnetic storms, can influence the muon flux measured at Earth's surface. These events are associated with coronal mass ejections (CMEs) and a stream of charged particles called the solar wind~\cite{Atri}. 

\medskip \noindent 
A Forbush decrease (FD) is detected when the intensity of cosmic radiation rapidly decreases, and is attributed to the deflection of some of the cosmic rays by the magnetic field associated with solar wind plasma following a CME. These decreases typically occur within hours, followed by a gradual flux recovery process lasting days until the cosmic ray radiation intensity returns to its initial level~\cite{Cane}.

\medskip \noindent 
In this paper, we quantify the fluctuations in muon flux using an array of scintillators at the Institute of Nuclear Sciences of UNAM. Additionally, we conduct a comprehensive analysis of geomagnetic data and identify multiple FDs occurring during the data-taking period. 

\section{Experimental Setup}
\begin{figure}[h!]
    \centering
    \includegraphics[width=\linewidth]{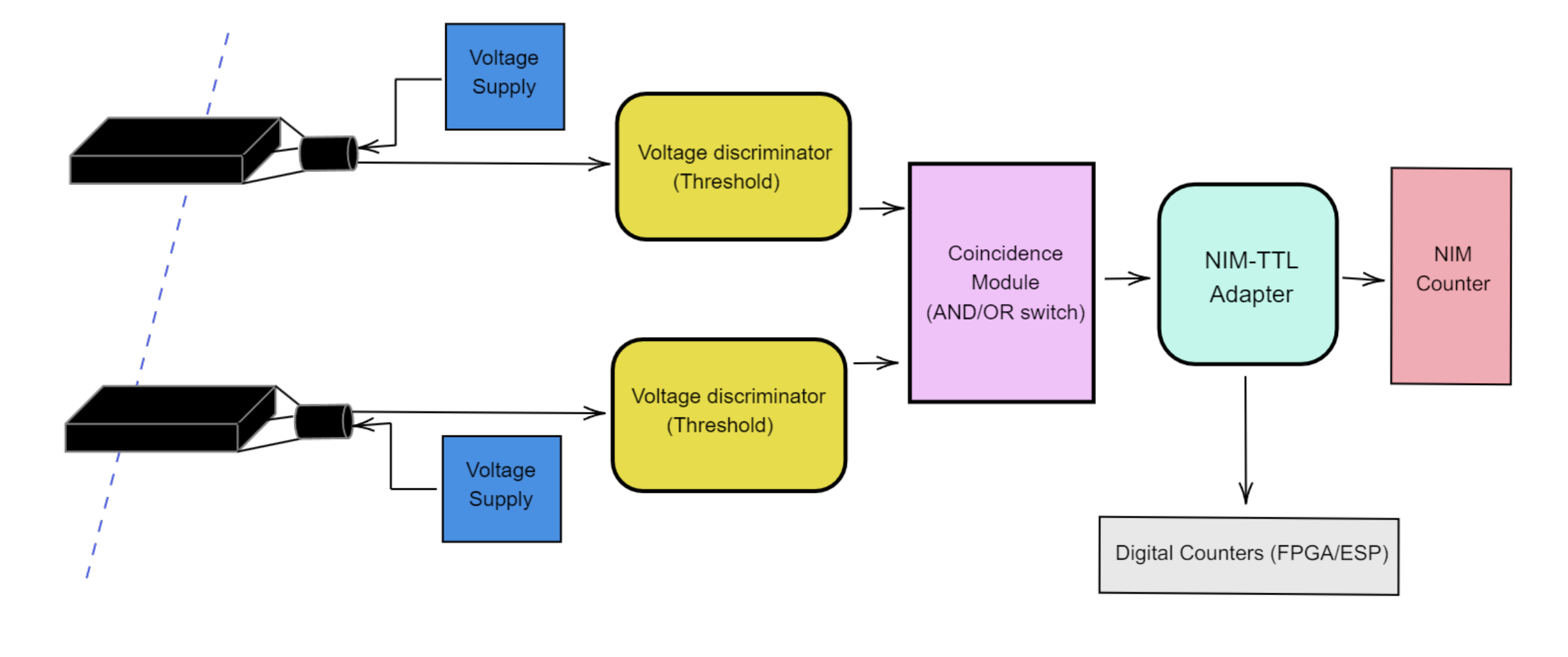}
    \caption{Diagram of the experimental setup, with a muon (dotted blue line) crossing the scintillators.}
    \label{electronica}
\end{figure}

\begin{figure}[h!]
    \centering
    \includegraphics[width=0.9\linewidth]{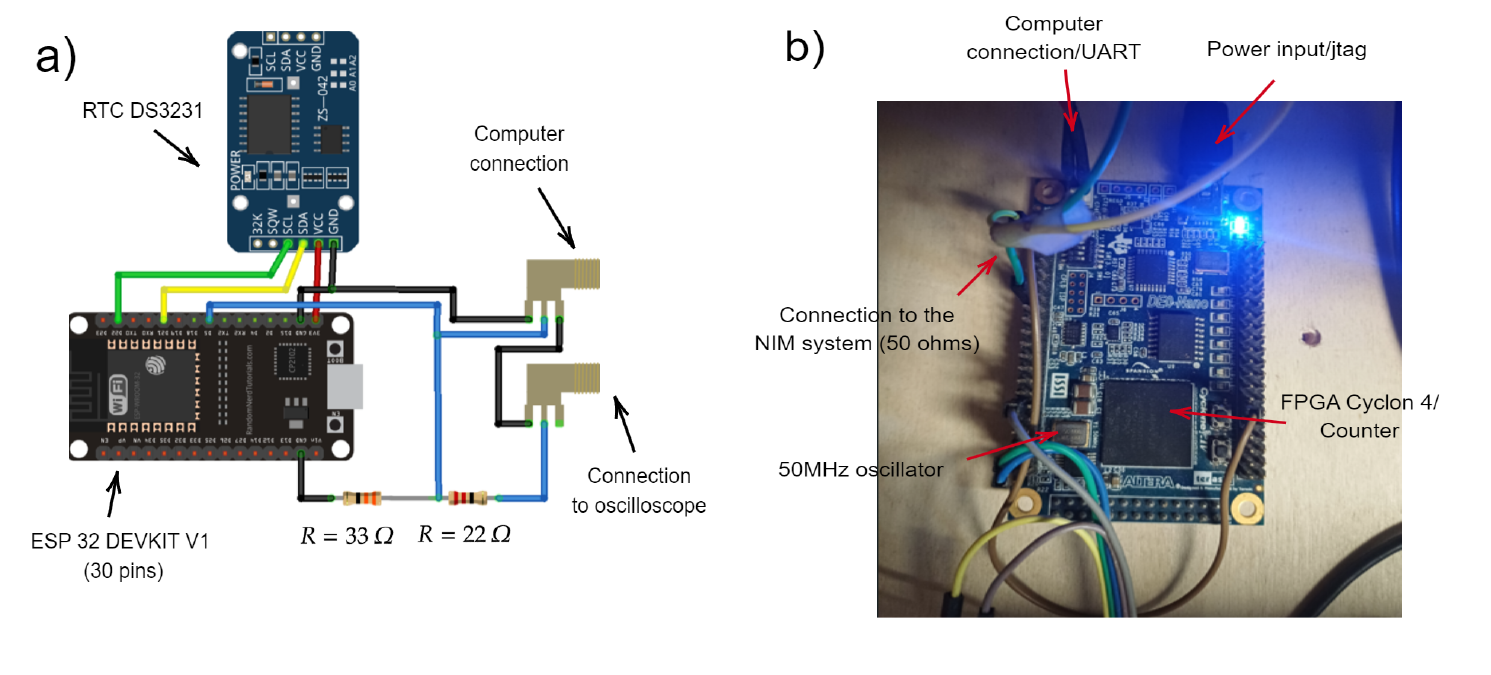}
    \caption{Schematic diagram of the digital counters: (a) the ESP32 counter and (b) the FPGA counter.}
    \label{counters}
\end{figure}

\noindent
The experimental setup, shown in Fig.~\ref{electronica}, consists of two plastic scintillators BNC-404, each measuring 50 cm x 1m. They are positioned with a separation of 10 cm between them and are individually coupled to a photomultiplier tube (PMT) Hamamatsu-R6231. We employ the electronic Nuclear Instrumentation Module (NIM), coupled to the PMTs. The Data Acquisition system involved a current signal from the PMTs to the discriminator module. Subsequently, the signal passed through a coincidence module and a NIM-TTL adapter before reaching the digital Field Programmable Gate Array (FPGA) and ESP32 DEVKIT microcontroller counters, shown in Fig.~\ref{counters}.

\noindent 
In order to determine the optimal operational voltage for the PMTs, we generated the plateau curve. This method involves monitoring the counting rate from the detector as a function of the applied voltage. A constant counting rate suggests that the detector's response has become less sensitive to variations in the PMT voltage, indicating an optimal operational region~\cite{Leo}. We performed a voltage sweep from 700 to 1500 V at various thresholds. Based on this analysis, the operational voltage for the PMTs was set at 1500 V, with a threshold adjusted to 10 mV.

\section{Results}

\medskip \noindent 
The measurement started on the 14th of September of 2023 at 23:00 UTC and ended on the 5th of December of 2023 at 23:00 UTC. Our muon detector is located at the Institute of Nuclear Sciences of UNAM in Mexico City, at coordinates 19.32$^{\circ}$N 99.18$^{\circ}$W, with an altitude of 2274 meters above see level. To reach our location, a muon must possess a minimum rigidity of 8.24 GV~\cite{OurRig}. Generally, cutoff values are primarily influenced by the geomagnetic latitude and vary from less than 1 GV near the geomagnetic poles to around 16 GV near the equator~\cite{Rig}.

\subsection{Measurement of the barometric coefficient}
In order to account for variations in air pressure, we analyze our dataset and derive the barometric coefficient $\beta$. The correlation between muon flux intensity and atmospheric pressure is characterized by~\cite{dorman}:
\begin{equation}
N = N_0\ e^{\beta\ \Delta P},
    \label{ec1}
\end{equation}
\noindent where $N$ represents the muon counts at pressure $P$, $\Delta P$ denotes the pressure difference with respect to the average, and the value of $N_0$ corresponds to the average value of muon counts.

\begin{figure}[h]
\centering
\includegraphics[width=\linewidth]{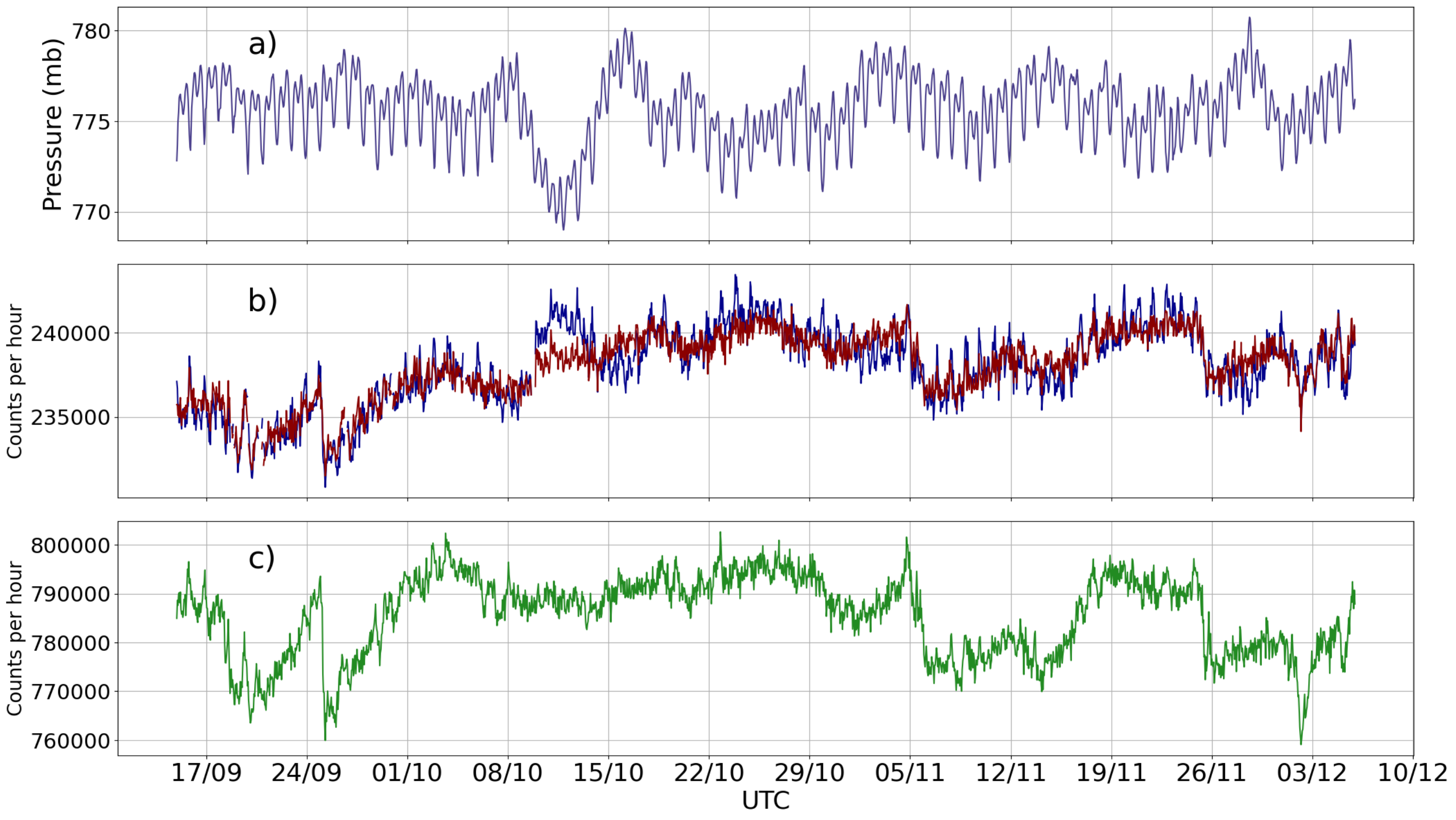}
\caption{(a) Pressure values in mb during the data-taking period. (b) Comparison between the initial muon counting rate (blue) and the rate after a correction for pressure is applied (red). (c) Hadronic counting rate, taken from~\cite{cr-unam}.}
\label{icn_geo}  
\end{figure}

\medskip \noindent 
To eliminate influences from solar or geomagnetic activity, we only analyzed data from geomagnetically quiet days, marked by low solar activity and no solar storms~\cite{omni,Sciesmex1}. The pressure was taken from the Cosmic Ray Observatory at UNAM~\cite{cr-unam}, shown in Fig. \ref{icn_geo}(a). We calculated a linear regression to determine $\beta$ by correlating variations in the muon rate with changes in atmospheric pressure. The result of our fit gives $\beta = -0.208$\% mb$^{-1}$. 

\medskip \noindent 
As expected, the counting rate for muons decreases with high atmospheric pressure values and increases when the pressure is low, illustrated in Fig. \ref{icn_geo}(b). Table~\ref{barometricos} presents a selection of barometric coefficients for muons observed in various detectors under different conditions. We observe that parameters such as altitude, cutoff rigidity, and latitude can influence the value of the barometric coefficient.

\subsection{Comparison with neutron flux measurements at UNAM}

The Cosmic Ray Observatory at UNAM~\cite{cr-unam} detects the hadronic component of secondary cosmic radiation by using a Neutron Supermonitor 6NM64. It has been in continuous operation since 1990 and the fluctuations in cosmic rays attributed to solar activity are monitored by the Mexican Space Weather Service~\cite{Sciesmex1}.

\medskip \noindent 
The comparison between the muon rate observed at ICN-UNAM after a correction for pressure is applied and the neutron rate measured by the Cosmic Ray Observatory is shown in Fig.~\ref{icn_geo}(b) and Fig. \ref{icn_geo}(c). We identify variations in the pressure-corrected muon flux that exhibit a pattern similar to those observed by a neutron detector

\begin{table}[h]
    \centering
    \begin{tabular}{ccccc} \hline\hline
        Detector & $\beta$ ($\%/mb$) & Altitude (m) & Cutoff (GV) & Reference  \\ \hline
        NGY (JAP)  & -0.114 & 77 & 11.5 &  \cite{mendoca} \\ 
        SMS (BRA)  & -0.119  & 48.8 & 9.3 &  \cite{mendoca} \\
        KWT (KUW)  & -0.124   & 19 & 13.8 & \cite{mendoca} \\
        HBT (AUS)  & -0.168 & 65 & 1.8 &  \cite{mendoca} \\ 
        MITO (SPA)  & -0.180  & 708 & 6.95 & \cite{mito} \\
        FINAPP (ITA) & -0.190 & 42 & 4.86 & \cite{Stevanato} \\
        KAAU (SAU)  & -0.240  & 27 & 14.8 & \cite{saudi2} \\
        CARPET (ARG)  & -0.440  & 2550 & 9.8 &  \cite{carpet} \\ 
\hline
        ICN (MEX)  & -0.208  & 2264 & 8.2 & this work \\
\hline\hline
    \end{tabular}
    \caption{Barometric coefficients for different muon detectors around the world.}
    \label{barometricos}
\end{table}

\subsection{Pressure-corrected muon flux measurement}

Given the geometry of our detector, we calculated the detector aperture~\cite{Piazzoli} to be equal to $A = 0.67$ m$^{2}$sr. Hence, the muon flux ($\mathcal{F}$) is determined according to the following equation:
\begin{equation}
    \mathcal{F} = \frac{N}{A \cdot t},
\end{equation}
\noindent 
where $t$ is the data-taking period, and $N$ is the number of muons observed during this period. The muon flux observed at ICN-UNAM is measured to be: 98.3 $\pm$ 2.6 (stat $\oplus$ syst) m$^{-2}$s$^{-1}$sr$^{-1}$, which agrees with a previous measurement of the flux at the same location of 101.2 $\pm$ 5.8 (stat $\oplus$ syst) \cite{master}. 

\subsection{Combined analysis of geomagnetic data and muon flux}

We investigate the variations in muon flux attributed to the arrival of the solar wind and CMEs from the Sun using the publicly available geomagnetic data~\cite{omni}. To estimate the potential impact of solar storms on the observed muon flux, we compared the variations with several geomagnetic parameters. 

\begin{figure}[h!]
\centering
\includegraphics[width=\linewidth]{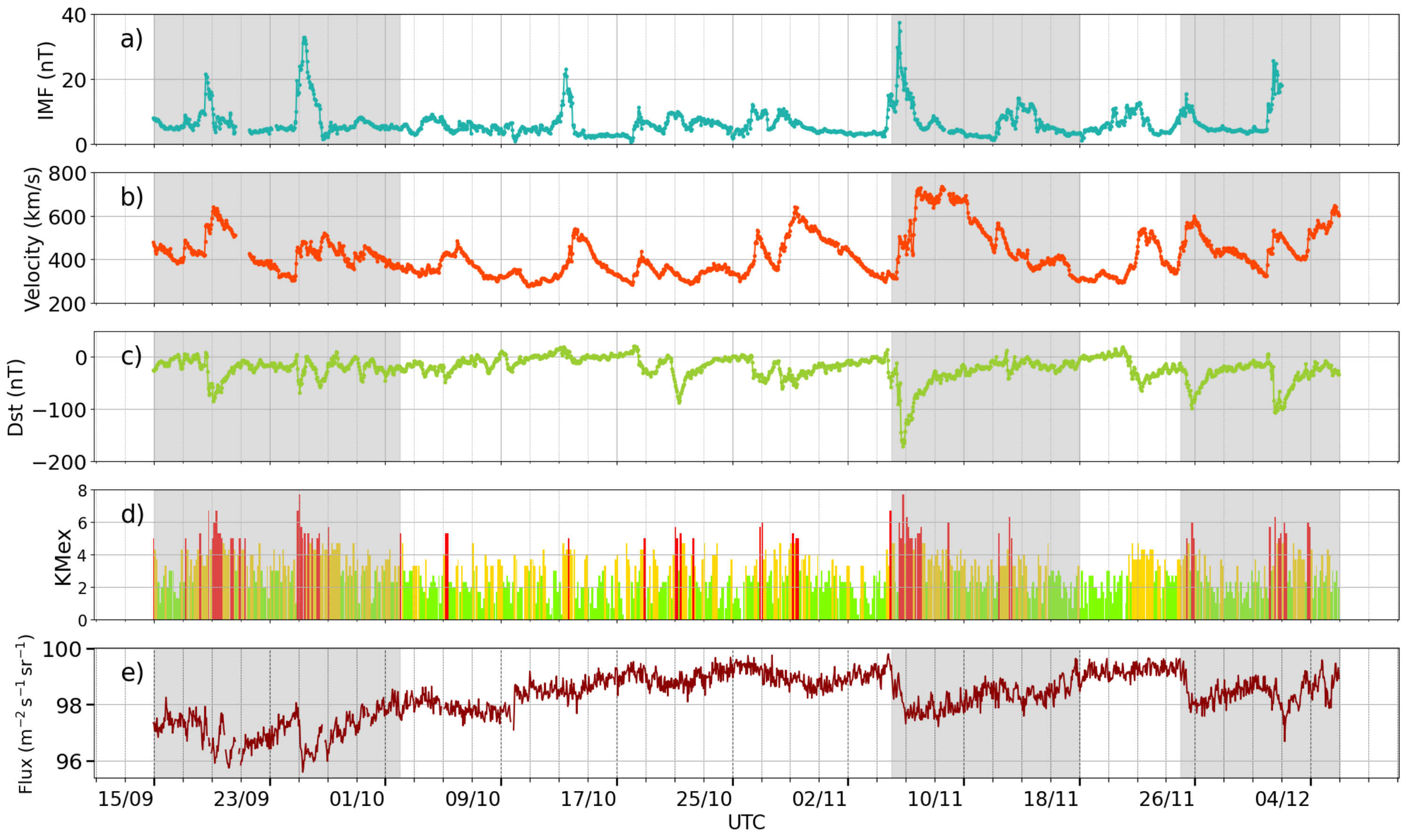}
\caption{
Geomagnetic data from September to December 2023: (a) interplanetary magnetic field, (b) solar wind speed , (c) Dst index and (d) Kmex index. The pressure-corrected muon flux as measured at ICN-UNAM is shown in (e). Gray zones indicate potential Forbush decreases.
}
\label{nasa}  
\end{figure}

\medskip \noindent 
Fig.~\ref{nasa} presents the geomagnetic data from September to December 2023 compared to the pressure-corrected muon flux as measured at ICN-UNAM. We analyze the interplanetary magnetic field, that influences the behavior of charged particles in the solar system, the disturbance storm time index (Dst), which quantifies the magnitude of geomagnetic storms, and the Kmex index~\cite{Sciesmex1}, a measure of the recorded variations in the geomagnetic field at Mexico's latitude.

\medskip \noindent 
We correlate the observed variations in the muon flux, shown in Fig.~\ref{nasa}(e), with three potential FD observed during the data-taking period, denoted as gray zones in Fig.~\ref{nasa}. According to geomagnetic data, during these periods, several CMEs arrived to Earth, leading to significant solar storms. As we observe from Fig.~\ref{nasa}(d), there are records of periods with Kmex values greater than 6, indicating G2 geomagnetic storms, as well as periods with Kmex greater than 7, indicating G3 geomagnetic storms~\cite{SWPC}.

\section{Conclusions} In this study, we report on the variations of atmospheric muon flux in Mexico City during the period from September to December 2023. The barometric coefficient $\beta$ is determined to be -0.21$\%$/mb. The observed muon flux is measured to be $98.3 \pm 2.6$ m$^{-2}$ s$^{-1}$ sr$^{-1}$, and it is consistent with other measurements at UNAM within the uncertainties. We identify three distinct variations in the pressure-corrected muon flux, classified as Forbush decreases. After analyzing publicly available geomagnetic data, we establish a correlation between these variations and large coronal mass ejections from the Sun.

\section*{Acknowledgments}
D.M.M. acknowledges the support received from the Mexican National Council of Humanities, Sciences, and Technologies (Conahcyt) as an assistant to an emeritus researcher (G.P.) within the National System of Researchers. L.S. acknowledges the support received from Conahcyt for the postdoctoral fellowship. We cordially thank Dr. Luis Xavier González Méndez and Dr. Ernesto Ortiz Fragoso for their valuable insights regarding the Cosmic Ray Observatory at UNAM and the data provided by the Mexican Space Weather Service.


\begin{thebibliography}{} 

\bibitem{Int1}  M. Potgieter, \href{https://doi.org/10.12942/lrsp-2013-3}{Living Rev. Sol. Phys. \textbf{10}, 3 (2013)}.

\bibitem{PGD} Particle Data Group, \href{https://lss.fnal.gov/archive/2022/pub/fermilab-pub-22-756.pdf}{Prog. Theor. Exp. Phys. 2022, 083C01 (2022)}.

\bibitem{Atri} D. Atri and A. Melott, \href{https://doi.org/10.1016/j.astropartphys.2013.03.001}{Astroparticle Physics \textbf{53}, 4 (2014)}

\bibitem{Cane} H. Cane, \href{https://doi.org/10.1023/A:1026532125747}{Space Sci. Rev. \textbf{93}, 55 (2000)}.

\bibitem{Leo}
W. Leo, \href{https://link.springer.com/book/10.1007/978-3-642-57920-2}{\textit{Techniques for Nuclear and Particle Physics Experiments}}, (Springer-Verlag, New York, 1994).

\bibitem{OurRig} B. Vargas-Cárdenas and J. F. Valdés-Galicia, \href{http://www.cbpf.br/~icrc2013/papers/icrc2013-0500.pdf}{ICRC2013 500 (2013)}.

\bibitem{Rig} D. Smart and M. Shea, \href{https://doi.org/10.1016/j.asr.2004.09.015}{Adv. in Space Research \textbf{36}, 2012 (2005)}.

\bibitem{dorman} L. Dorman, \href{https://link.springer.com/book/10.1007/978-1-4020-2113-8}{\textit{Cosmic Rays in the Earth’s Atmosphere and Underground}}, (Springer Science, New York, 2004)

\bibitem{omni}
OMNIWeb Plus data, NASA, \href{https://omniweb.gsfc.nasa.gov}{https://omniweb.gsfc.nasa.gov}.

\bibitem{Sciesmex1}
Mexican Space Weather Service, \href{https://www.sciesmex.unam.mx}{https://www.sciesmex.unam.mx}.

\bibitem{cr-unam}
Cosmic Ray Observatory at UNAM, Mexico, \href{http://www.cosmicrays.unam.mx}{http://www.cosmicrays.unam.mx}.

\bibitem{mendoca} R. de Mendonça \textit{et al}., \href{https://iopscience.iop.org/article/10.3847/0004-637X/830/2/88}{Astrophysical Journal \textbf{830}, 88 (2016)}.

\bibitem{mito} S. Ayuso \textit{et al}., \href{https://doi.org/10.1051/swsc/2020079}{J. Space Weather Space Clim. \textbf{11}, 13 (2021)}.

\bibitem{Stevanato} L. Stevanato \textit{et al}., \href{https://doi.org/10.1029/2021GL095383}{Geophys. Res. Lett. \textbf{49}, 6 (2022)}.

\bibitem{saudi2} A. Maghrabi \textit{et al}., \href{https://doi.org/10.4236/ijaa.2023.133014}{Inter. J.  of Astronomy and Astrophysics \textbf{13}, 236 (2023)}.

\bibitem{carpet} R. de Mendonça \textit{et al}., \href{https://doi.org/10.1029/2012JA018026}{J. Geophys. Res. Space Physics \textbf{118}, 1403 (2013)}.

\bibitem{Piazzoli} B. D'Ettorre Piazzoli \textit{et al}., \href{https://doi.org/10.1016/0029-554X(76)90170-1}{Nucl. Instrum. Meth. \textbf{135}, 223 (1976)}.

\bibitem{master} B. Olmos and A. Aguilar-Arevalo, \href{https://doi.org/10.1016/j.nima.2020.164870}{Nucl. Instrum. Meth. \textbf{987}, 164870 (2021)}.

\bibitem{SWPC} NASA Space Weather Prediction Center,  \href{https://www.swpc.noaa.gov/}{https://www.swpc.noaa.gov/}.

\end{thebibliography}
\end{document}